\documentclass[prb,aps,twocolumn,amsmath,amssymb]{revtex4}

\usepackage{epsfig}% Include figure files
\usepackage{graphicx}% Include figure files
\usepackage{dcolumn}% Align table columns on decimal point
\usepackage{bm}% bold math

\begin{document}

\title{One-electron scattering rate and normal-state linear-$T$ resistivity of the cuprates}
\author{J. M. P. Carmelo}
\affiliation{GCEP-Centre of Physics, 
University of Minho, Campus Gualtar, P-4710-057 Braga, Portugal.}
%\affiliation{Department of Physics,
%Massachusetts Institute of Technology, Cambridge, MA 02139-4307, USA}

%%%%%%%%%%%%%%%%%%%%%%%%%%%%%%%%%%%%%%%%%%%%%%%%%%%%%%%%%%%%%%%%%%%%%%%%%%
%                              abstract                                  %
%%%%%%%%%%%%%%%%%%%%%%%%%%%%%%%%%%%%%%%%%%%%%%%%%%%%%%%%%%%%%%%%%%%%%%%%%%

\begin{abstract}
Here we use a description of the electronic correlations contained in the Hubbard model 
on the square-lattice perturbed by very weak three-dimensional uniaxial anisotropy 
in terms of the residual interactions of charge $c$ fermions and spin-neutral composite two-spinon 
$s1$ fermions. Excellent quantitative agreement with  
the anisotropic linear-$\omega$ one-electron scattering rate and normal-state linear-$T$ resistivity 
observed in experiments on hole-doped cuprates 
with critical concentrations $x_c\approx 0.05$ and $x_*\approx 0.27$ is achieved.
Our results provide strong evidence that the normal-state linear-$T$ resistivity is a 
manifestation of low-temperature scale-invariant physics.
\end{abstract}

\maketitle
%%%%%%%%%%%%%%%%%%%%%%%%%%%%%%%%%%%%%%%%%%%%%%%%%%%%%%%%%%%%%%%%%%%%%%%%%%
%                              body of paper                             
%%%%%%%%%%%%%%%%%%%%%%%%%%%%%%%%%%%%%%%%%%%%%%%%%%%%%%%%%%%%%%%%%%%%%%%%%%
%\tableofcontents

%\date{January 2010}

\section{Introduction}

There is evidence of quantum critical behaviour in the 
cuprate superconductors \cite{LSCO-1/lam-x,critical,critical-2,Resistivity,QPT}. Its 
physical nature and the mechanism underlying the pairing state 
\cite{ARPES-review,2D-MIT,duality,two-gaps} are not well understood.  
Here we use a description of the electronic correlations contained in the Hubbard model 
on the square-lattice \cite{ARPES-review,2D-MIT,companion2} perturbed by very weak 
three-dimensional (3D) uniaxial anisotropy 
in terms of the residual interactions of charge $c$ fermions and spin-neutral composite two-spinon 
$s1$ fermions \cite{companion2,cuprates,cuprates0}. The interplay of the $c$ Fermi line isotropy
with the $s1$ band boundary line anisotropy plays a major
role in the one-electron scattering properties. Excellent quantitative agreement with  
the anisotropic linear-$\omega$ one-electron scattering rate and normal-state linear-$T$ resistivity 
observed in experiments on hole-doped cuprates 
with critical concentrations $x_c\approx 0.05$ and $x_*\approx 0.27$ is achieved.
Our results provide strong evidence that the normal-state linear-$T$ resistivity is a 
manifestation of low-temperature scale-invariant physics near a hole 
concentration $x_{c2}\approx 0.20-0.22$.

Recent experiments on the hole-doped cuprate superconductors
\cite{critical-2,Resistivity,two-gaps,Yu-09,coex,mom,iti,k-r-spaces,Delta-no-super,spiral-order,PSI-ANI-07,two-gap-Bi,Hm,LSCO-ARPES-peaks,two-gap-Tanaka,
two-gap-Tacon,scattering-rate} impose new severe constraints on the mechanisms responsible
for their unusual properties. The above toy model provides the simplest realistic  
description of the role of correlations effects in these properties. 
It involves the effective transfer integrals $t$ and $t_{\perp}\ll t$ and 
effective on-site repulsion $U$. 

An extended presentation of the results of this manuscript, including 
needed supplementary information, can be found in Ref. \cite{cuprates}. 

\section{The virtual-electron pair quantum liquid}

The virtual-electron 
pair quantum liquid (VEPQL) \cite{cuprates0} describes 
that toy model electronic correlations in terms of residual
$c$ - $s1$ fermion interactions. Alike the Fermi-liquid 
quasi-particle momenta \cite{cuprates,Pines}, those of the $c$ and $s1$ fermions 
are close to good quantum numbers \cite{companion2,cuprates0}. 

The results of Ref. \cite{cuprates0} provide strong evidence that 
for hole concentrations $x\in (x_c,x_*)$ the VEPQL short-range spin order coexists 
with a long-range $d$-wave superconducting order
consistent with unconventional superconductivity being mediated by magnetic fluctuations \cite{Yu-09}.
The  {\it virtual-electron pair} configurations involve one zero-momentum $c$ fermion pair
of charge $-2e$ and one spin-singlet two-spin-$1/2$ spinon $s1$ fermion. 
The VEPQL predictions achieve an excellent agreement with 
the cuprates universal properties \cite{cuprates0} and consistency with the coexisting two-gap scenario 
\cite{two-gaps,Delta-no-super,two-gap-Tanaka,two-gap-Tacon}:
%%%%%%%%%%%%%%%%%%%%%%%%%%%%%%%%%%%%%%%%%%%%%%%%%%%%%%%%%%%%%%%%%%%%%%%%%%%%%
% Figure
%%%%%%%%%%%%%%%
\begin{figure}
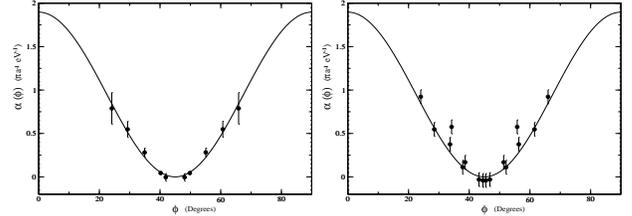

\includegraphics[width=4.0cm,height=2.86cm]{alphaI-new-08.eps}
\includegraphics[width=4.0cm,height=2.86cm]{alphaII-new-08.eps}
\caption{{\bf The theoretical and LSCO scattering-rate linear-$\omega$ coefficient.} 
The theoretical coefficient $\alpha (\phi)$ given in Ref. \cite{cuprates}
(solid line) for $x=0.145$
and the LSCO parameters of the Table I
together with the experimental points for the corresponding
coefficients $[\alpha_I (\phi)-\alpha_I (\pi/4)]$ and
$[\alpha_{II} (\phi)-\alpha_{II} (\pi/4)]$ 
of Fig. 4 (c) of Ref. \cite{PSI-ANI-07}.}
\label{fig1} 
\end{figure}
A dome-like superconducting energy scale $2\vert\Omega\vert$ and 
pseudogap $2\vert\Delta\vert$, over the whole dome $x\in (x_c,x_*)$. Those are 
the maximum magnitudes of the superconducting virtual-electron pairing
energy and spin-singlet spinon pairing energy, respectively. 

In the ground state there is no one-to-one correspondence between a $c$ fermion pair
and a two-spinon $s1$ fermion: The strong effective coupling of $c$ fermion pairs whose hole
momenta $\vec{q}^{\,h}$ and $-\vec{q}^{\,h}$ belong to an approximately circular
$c-sc$ line centered at $-\vec{\pi}=-[\pi,\pi]$, of radius 
$q^h=\vert\vec{q}^{\,h}\vert\in (q^h_{Fc},q^h_{ec})$ and $c$ fermion 
energy $\vert\epsilon_c (q^h)\vert \in (0,W_{ec})$,
results from interactions with well-defined $s1$ fermions. Those have 
$s1$ band momentum $\vec{q}$ belonging to a corresponding $s1-sc$ line arc centered 
at $\vec{0}=[0,0]$. Here $\vert\epsilon_c (q^h_{Fc})\vert=0$ and  
$\vert\epsilon_c (q^h_{ec})\vert=W_{ec}$. For Fermi angles $\phi\in (0,\pi/2)$
a $s1-sc$ line arc can be labelled either by its nodal momentum absolute
value $q^N_{arc} = q^N_{arc} (q^h)\in (q^N_{ec},q^N_{Bs1})$ or angular width
$2\phi_{arc}\in (0,\pi/2)$ given in Ref. \cite{cuprates}. $q^N_{Bs1}=q^N_{Bs1}(q^h_{Fc})$ 
is the $s1$ band nodal boundary-line momentum \cite{companion2,cuprates0}, 
$q^h_{Fc}\approx \sqrt{x\,\pi}\,2$ and $q^N_{ec}=q^N_{ec}(q^h_{ec})$ and
$q^h_{ec}$ are given in that reference. The energy needed for the
$c$ fermion strong effective coupling is supplied by the short-range spin correlations through the $c$ - $s1$
fermion interactions within each virtual-electron pair configuration. Virtual-electron pairs
exist in virtual intermediate states of pair-breaking one-electron and spin excitations.
Provided that the $c$ fermion effective coupling is strong,
virtual-electron pair breaking under one-electron removal excitations gives rise to sharp-feature-line arcs 
centered at momenta $\pm\vec{\pi}=\pm [\pi,\pi]$
in one-to-one correspondence to the $s1-sc$-line arcs of its $s1$ fermion. Such
sharp-feature-line arcs have angular range $\phi\in (\pi/4 -\phi_{arc},\pi/4 +\phi_{arc})$,
energy $E \approx 2W_{ec} (1-\sin 2\phi_{arc})$ and well-defined hole momentum 
$\vec{k}^{\,h} = \vec{k} + \vec{\pi}$ and
thus exist only for $E < E_1 (\phi) = 2W_{ec}(1-\vert\cos 2\phi\vert)$. 
The macroscopic condensate refers to zero-momentum $c$ fermion pairs  
whose phases $\theta=\theta_0 +\theta_1$ line up. The fluctuations of $\theta_0$ and $\theta_1$ 
become large for $x\rightarrow x_c$ and $x\rightarrow x_*$, respectively. The dome $x$
dependence of the critical temperature $T_c$ given in Refs. \cite{cuprates,cuprates0} 
is fully determined by the interplay of such fluctuations. 
A pseudogap state with short-range spin order and virtual-electron pair configurations without phase coherence occurs 
for temperatures $T\in (T_c,T^*)$ where $T^*$ is the pseudgap temperature also given in these references.
At $T=0$ a normal state emerges by application of a magnetic field aligned perpendicular to the planes of magnitude 
$H\in (H_0,H_{c2})$ for $x\in (x_0,x_{c2})$ and $H\in (H_0,H^*)$ for $x\in (x_{c2},x_*)$. 
The fields $H_0$, $H_{c2}$ and $H^*$ are given in Refs. \cite{cuprates,cuprates0}.

\section{The one-electron scattering rate and normal-state linear-$T$ resistivity of the hole-doped cuprates}

The main goals of this paper are: i) The study of the one-electron scattering rate and normal-state $T$-dependent 
resistivity within the VEPQL; ii) Contributing to the further understanding of the role of scale-invariant physics in the unusual
scattering properties of the hole-doped cuprates. The parameters appropriate to YBa$_2$Cu$_3$O$_{6+\delta}$ (YBCO 123)
and La$_{2-x}$Sr$_x$CuO$_4$ (LSCO) found in Ref. \cite{cuprates0} are given in Table I.
\begin{table}
\begin{tabular}{|c|c|c|c|c|} \hline
& $U/4t$ & $t$ & $\Delta_0$ & $x_c$ \\
system & & (meV) & (meV) & \\
\hline
YBCO 123 & $1.525$ & $295$ & $84$ & $0.05$ \\
\hline
LSCO & $1.525$ & $295$ & $42$ & $0.05$ \\
\hline
\end{tabular}
\caption{The magnitudes of four basic parameters appropriate to YBCO 123 and LSCO within the VEPQL scheme of Ref. [11]. 
The magnitude of the ratio $U/4t$ given here is set so that the $T=0$ dome upper critical hole concentration 
reads $x_*=0.27$. Those of the effective transfer integral $t$ and energy parameter $\Delta_0$ 
control the magnitudes of the VEPQL energy scales. The magnitudes of the suppression coefficient $\gamma_d^{min}=0.95$ 
for YBCO 123 and $\gamma_d^{min}=0.82$ for LSCO reached at $x=x_{op}=0.16$
account for the small effects of superfluid density anisotropy and intrinsic disorder, respectively, on the critical 
temperature $T_c$ of Eq. (\ref{T-c}). They do not play an important role in the quantities
studied in the letter. Moreover, the magnitudes $\varepsilon^2=6.2\times 10^{-3}$ for 
YBCO 123 and $\varepsilon^2=3.4\times 10^{-4}$ for LSCO of the very small 3D uniaxial anisotropy coefficient 
$\varepsilon^2\propto t_{\perp}/t$ defined in Ref. \cite{cuprates0} are set so that the $T=0$ lower dome critical hole 
concentration reads $x_c\approx 0.05$, as given in this table.}
\label{table1}
\end{table} 

Our results refer to a range $x\in (x_A,x_{c2})$ 
for which $V_{Bs1}^{\Delta}/V_{Fc}\ll 1$ where the $s1$ boundary line and $c$ Fermi velocities are
given in Refs. \cite{cuprates,cuprates0} and $x_A\approx x_*/2 =0.135$. 
For $x\in (x_{c2},x_*)$ that condition is also fulfilled but there emerge competing scattering processes difficult to describe in 
terms of $c$ - $s1$ fermion interactions. Consistently with the validity of our scheme, in 
Ref. \cite{cuprates} it is shown that the VEPQL
predictions agree quantitatively with the distribution of the experimental LSCO sharp photoemission 
spectral features. As predicted, they occur for energies $E (\phi)< E_1 (\phi)$
and the corresponding energy, nodal-momentum and sharp-feature line arcs angular ranges
agree with the theoretical magnitudes. Such an analysis reveals clear 
experimental spectral signatures of the VEPQL virtual-electron pairing mechanism.
A Fermi's golden rule in terms of the $c$ - $s1$ fermion interactions is then used in Ref. \cite{cuprates} to 
calculate for the parameters of Table I, small $\hbar\omega$ and both the normal and superconducting 
states the one-electron inverse lifetime $\hbar/\tau_{el}\approx \hbar\omega\,\pi\alpha_{\tau_{el}}$ 
and scattering rate $\Gamma (\phi,\omega) =1/[\tau_{el} \,V_F]\approx \hbar\omega\,\pi\alpha$. 
Here $V_F$ is the Fermi velocity, $\alpha_{\tau_{el}} = [\pi/16 \sqrt{x\,x_{op}}](\cos 2\phi)^2$ and 
$\alpha = (\cos 2\phi)^2/(x\,64x_*\sqrt{\pi x_{op}}\,t)$ where $x_{op}=(x_*+x_c)/2=0.16$. (We use units
of lattice constant $a=1$.) As discussed in Ref. \cite{cuprates}, such small-$\hbar\omega$ expressions 
are expected to remain valid for approximately $\hbar\omega <E_1 (\phi)$.
The factor $(\cos 2\phi)^2$ also appears in the anisotropic component
of the scattering rate studied in Ref. \cite{scattering-rate} for hole concentrations $x>x_{c2}$.
For $x=0.145$ the use of the Table I LSCO parameters leads to the theoretical 
coefficient $\alpha (\phi)$ plotted in Fig. \ref{fig1} (solid line) 
together with the experimental points
of Fig. 4 (c) of Ref. \cite{PSI-ANI-07} for the
coefficients $[\alpha_I (\phi)-\alpha_I (\pi/4)]$ and
$[\alpha_{II} (\phi)-\alpha_{II} (\pi/4)]$.
(The very small $\alpha_{I} (\pi/4)$ magnitude 
and small $\alpha_{II} (\pi/4)$ magnitude are related 
to processes that are not contained in the VEPQL.)
An excellent quantitative agreement is obtained  
between $\alpha (\phi)$ and the experimental points of both
$[\alpha_I (\phi)-\alpha_I (\pi/4)]$ and $[\alpha_{II} (\phi)-\alpha_{II} (\pi/4)]$.
%%%%%%%%%%%%%%%%%%%%%%%%%%%%%%%%%%%%%%%%%%%%%%%%%%%%%%%%%%%%%%%%%%%%%%%%%%%%%
% Figure
%%%%%%%%%%%%%%%
\begin{figure}
\includegraphics[width=3.75cm,height=3.75cm]{resist-L-new-sirius.eps}
\includegraphics[width=3.75cm,height=3.75cm]{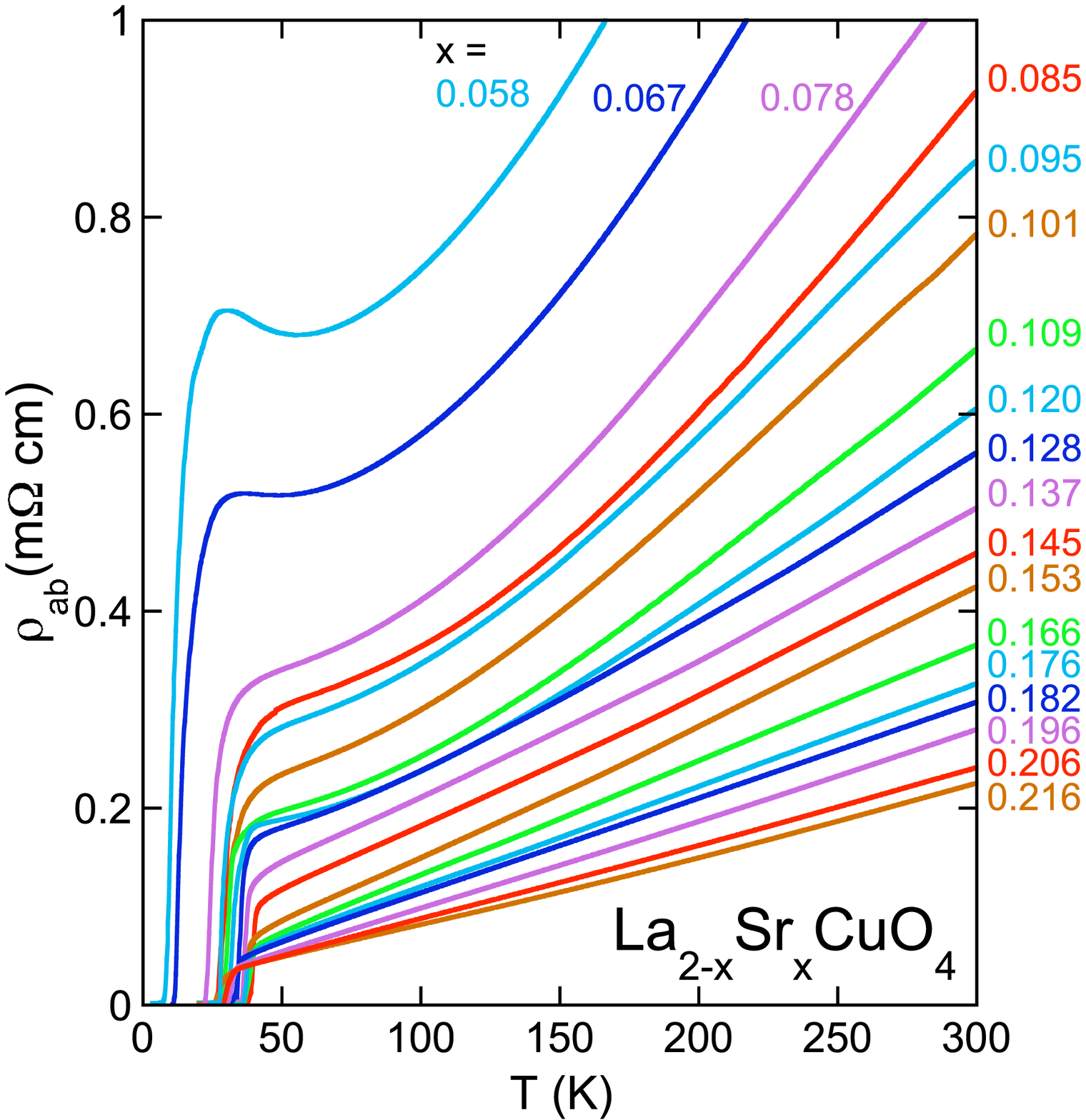}
\caption{{\bf Theoretical and LSCO linear-$T$ resistivity.}
The $T$ dependence of the resistivity
$\rho (T,0)=\theta (T-T_c)\,\rho (T)$ with
$\rho (T)$ given in Eq. (\ref{rho-F}) for $x\in (x_A,x_{c2})$
where $x_A\approx 0.135$ and $x_{c2}\approx 0.20-0.22$ and the parameter 
magnitudes of the Table I for LSCO and 
corresponding experimental curves.
Experimental curves figure from Ref. \cite{L-resistivity}.}
\label{fig2} 
\end{figure}

In the following we provide strong evidence from agreement between theory
and experiments that the linear-$T$ resistivity is indeed 
a manifestation of normal-state scale-invariant physics. 
This requires that the $T$-dependence of the inverse relaxation lifetime 
derived for finite magnetic field, $x\in (x_A,x_{c2})$ and
$\hbar\omega\ll \pi k_B\,T$ by replacing $\hbar\omega$ by $\pi k_B\,T$ in the  
one-electron inverse lifetime $1/\tau_{el}$ and averaging over the Fermi line
leads to the observed low-$T$ resistivity. The use in the Appendix A of that
procedure leads to the following resistivity expression,
\begin{equation}
\rho (T) \approx \left({\hbar \,d_{\parallel}\over t e^2}\right)\left({\pi\over 4}\right)^3{x_*\over x^{3/2}\sqrt{x_{op}}}\,\pi k_B T  \, .
\label{rho-F}
\end{equation}
Consistency with the above $\hbar/\tau_{el}$ expression validity range
$\hbar\omega <E_1 (\phi)$ implies that the behaviour (\ref{rho-F}) remains dominant 
in the normal-state range $T\in (0,T_1)$. Here, 
\begin{equation}
T_1\approx {2\over\pi}\int_{0}^{\pi/2}d\phi {E_1(\phi)\over k_B}= {(2\pi-4)W_{ec}\over \pi^2 k_B} \, .
\label{T-1}
\end{equation}
At $x=0.16$ use of the parameters of the Table I gives
$T_1\approx 554$\,K for LSCO and $T_1\approx 1107$\,K for 
YBCO 123. Extrapolation of expression (\ref{rho-F}) to $H=0$ 
leads to $\rho (T,0)\approx \theta (T-T_c) \rho (T)$ for $T<T_1$. Here
the critical low-$T$ resistivity behaviour (\ref{rho-F}) is masked by the 
onset of superconductivity at $T=T_c$. 

\begin{figure}
\includegraphics[width=3.75cm,height=3.75cm]{resist-new-new.eps}
\includegraphics[width=3.75cm,height=3.75cm]{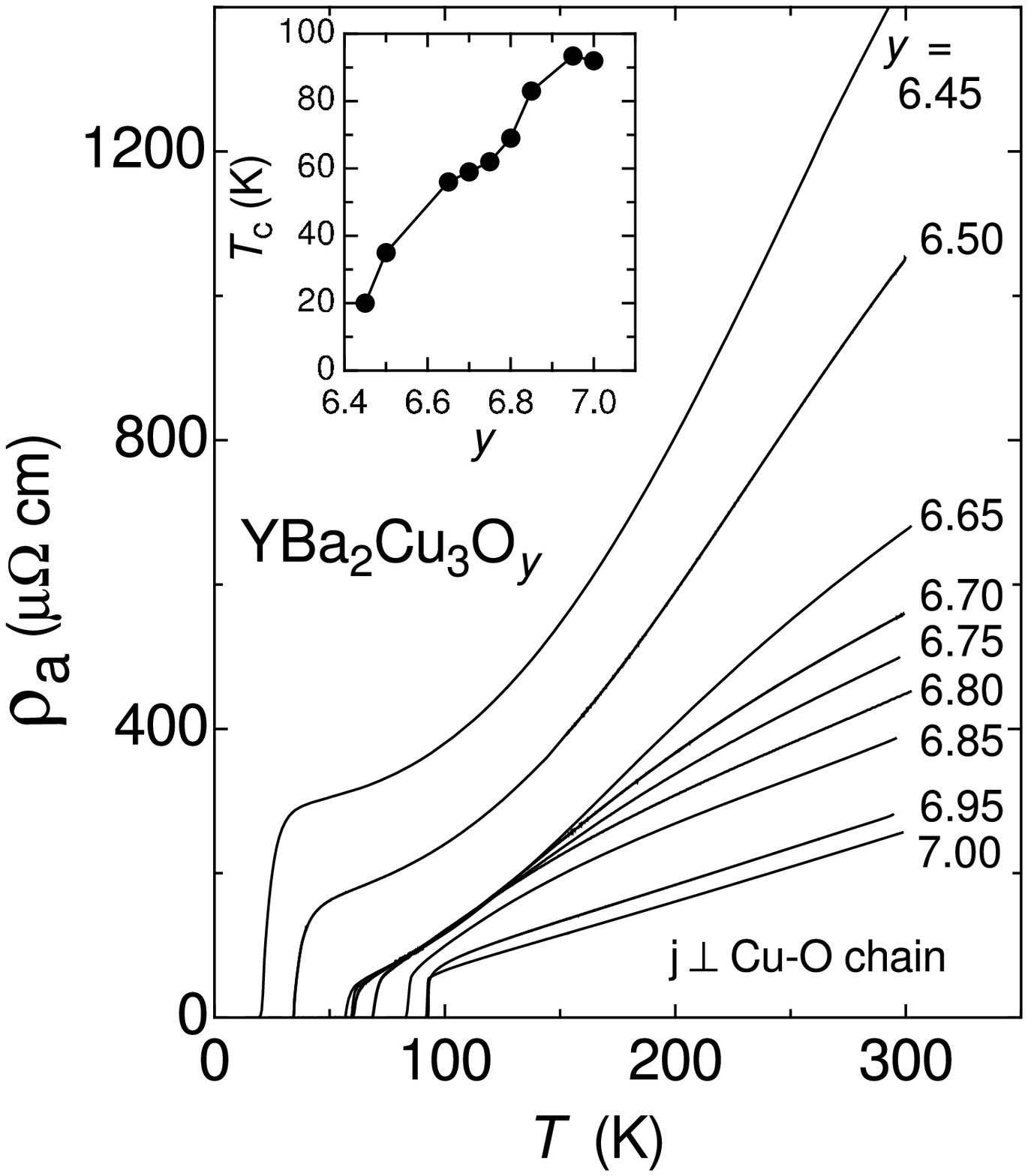}
\caption{{\bf Theoretical and YBCO 123 linear-$T$ resistivity.}
The $T$ dependence of the resistivity
$\rho_a (T,0)=\theta (T-T_c)\,\rho (T)$ where $\rho (T)$
is given in (\ref{rho-F}) for the parameter 
magnitudes of the Table I for YBCO 123 and corresponding
experimental curves of Ref. \cite{Y-resistivity}
for a set of $y$ values. The oxygen content $y-6$ is obtained from
$x$ by use of Fig. 4 (a) of Ref. \cite{Tc-1/8}. The theoretical $T_c$ is 
larger at $y=6.95$ than for $y=6.85,7.00$. This is alike in the inset of the 
second figure. Experimental curves figure from Ref. \cite{Y-resistivity}.}
\label{fig3} 
\end{figure}

We now compare our theoretical linear-$T$ resistivity with that of LSCO \cite{L-resistivity} and
YBCO 123 \cite{Y-resistivity} for $H=0$ and $T$ up to $300$\,K.
Transport in the $b$ direction has for YBCO 123
contributions from the CuO chains, which render our results unsuitable. In turn,
$\rho_a (T,0)\approx \theta (T-T_c) \rho (T)$ at $H=0$ for the $a$ direction. 
$\rho (T,0)$ and $\rho_a (T,0)$ are plotted in Figs. \ref{fig2} and
\ref{fig3} for the Table I parameters for LSCO and YBCO 123, respectively. $x$  
is between $x\approx x_A\approx 0.135$ and $x_{c2}\approx 0.20-22$ for the 
LSCO theoretical lines of Fig. \ref{fig2}. Fig. \ref{fig3} for YBCO 123 refers to three $x$ values near 
$x_{op}$, expressed in terms of the oxygen content.
Comparison of the theoretical curves of Fig. \ref{fig2}
with the LSCO resistivity curves of Ref. \cite{L-resistivity} also shown
in the figure confirms an excellent quantitative agreement
between theory and experiments for the present range $x\in (x_A,x_{c2})$. 
In turn, for YBCO 123 our scheme provides a good quantitative description of the experimental curves
near $x_{op}$, for $y=6.95,7.00$. The $y=6.85$ experimental curve of Ref. \cite{Y-resistivity} 
already deviates from the linear-$T$ behaviour. The corresponding effective hole
concentration $x_A\approx 0.15$ 
is for that material consistent with the inequality $x_A\in (0.12,0.15)$ found in Ref. \cite{cuprates}. 
In turn, for $x>x_{c2}$ a competing scattering channel 
emerges, leading to an additional $T^2$-quadratic resistivity contribution \cite{critical-2,Resistivity}. 
%%%%%%%%%%%%%%%%%%%%%%%%%%%%%%%%%%%%%%%%%%%%%%%%%%%%%%%%%%%%%%%%%%%%%%%%%%%%%
% Figure
%%%%%%%%%%%%%%%

\section{Concluding Remarks}

That the dependence on the Fermi angle  $\phi\in (0,\pi/2)$ of the scattering-rate
coefficient $\alpha=\alpha_{\tau_{el}}/\hbar V_F$ 
associated with that of the inverse lifetime $\hbar/\tau_{el}\approx \hbar\omega\,\pi\alpha_{\tau_{el}}$ agrees with the experimental 
points of Fig. \ref{fig1} seems to confirm the evidence provided in
Refs. \cite{companion2,cuprates0} that the VEPQL may contain some of
the main mechanisms behind the unusual properties of the hole-doped cuprates.
That in addition the inverse relaxation lifetime $1/\tau$ $T$-dependence 
obtained for $\hbar\omega\ll \pi k_B\,T$ and $x\in (x_A,x_{c2})$
by merely replacing $\hbar\omega$ by $\pi k_B\,T$ in $1/\tau_{el}$ 
and averaging over the Fermi line leads to astonishing quantitative agreement with 
the resistivity experimental lines is a stronger surprising result. 
Consistently, for $\hbar\omega\ll \pi k_B\,T$ and $x\in (x_A,x_{c2})$ the system exhibits dynamics 
characterized by the relaxation time $\tau = \hbar A/\pi k_B\,T$ of Eq. (\ref{1-over-tau})
of the Appendix A, where $A = [32\sqrt{x\,x_{op}}/\pi^2]\approx 1$ for $x> x_A$. 
This second stronger result provides clear evidence
of normal-state scale-invariant physics. It may follow from beyond mean-field theory the $T=0$ 
line $H_{c2}(x)$, plotted in Fig. S1 for $x\in (x_0,x_{c2})$, referring to a true quantum phase transition.
Such a transition could occur between a state with long-range spin order 
regulated by monopoles and antimonopoles for $H>H_{c2}$ and a vortex liquid  
by vortices and antivortices for $H<H_{c2}$. 
That would be a generalization for $H_{c2}>0$ of the quantum phase transition speculated to occur 
at $x_{c2}\approx x_0$ and $H_{c2}\approx 0$ in Ref. \cite{duality}. The field $H^{*}$ 
marks a change from a short-range spin order to a disordered state and thus refers to a crossover.
Hence the normal-state scale invariance occurring for $x\in (x_A,x_{c2})$ could result from 
the hole concentration $x_{c2}\approx 0.2$, where the lines of Fig. S1
associated with the fields $H_{c2} (x)$ and $H^* (x)$ meet, referring to a quantum critical point
\cite{LSCO-1/lam-x,critical}. Such a quantum critical point may prevent the competing 
$T^2$-quadratic resistivity contribution to strengthen below $x=x_{c2}$.

%%%%%%%%%%%%%%%%%%%%%%%%%%%%%%%%%%%%%%%%%%%%%%%%%%%%%%%%%%%%%%%%%%%%%%%%%%
\begin{acknowledgments}
I thank M. A. N. Ara\'ujo, J. Chang, A. Damascelli, R. Dias, P. Horsch, S. Komiya, 
J. Mesot, C. Panagopoulos, T. C. Ribeiro, P. D. Sacramento, M. J. Sampaio, M. Shi and 
Z. Te${\rm\check{s}}$anovi\'c for discussions
and the support of the ESF Science Program INSTANS and grant PTDC/FIS/64926/2006.
\end{acknowledgments}
%%%%%%%%%%%%%%%%%%%%%%%%%%%%%%%%%%%%%%%%%%%%%%%%%%%%%%%%%%%%%%%%%%%%%%%%%%
%%%%%%%%%%%%%%%%%%%%%%%%%%%%%%%%%%%%%%%%%%%%%%%%%%%%%%%%%%%%%%%%%%%%%%%%%%
\appendix

\section{Methods used in our studies}

Our general method relies on the use of a suitable description in terms of $c$ and
$s1$ fermions whose momenta are close to good quantum numbers
of the square-lattice quantum liquids studied in Refs. \cite{companion2,cuprates0}.
The low-$T$ resistivity (\ref{rho-F}) is accessed for the normal state reached by 
applying a magnetic field perpendicular to the planes, which
remains unaltered down to $T=0$, as in the cuprates \cite{critical-2}.
The field serves merely to remove superconductivity and achieve the $H$-independent 
term $\rho (T)$ of, 
\begin{equation}
\rho (T,H) = \rho (T)+\delta\rho (T,H) \, ,
\label{rho-T-TH}
\end{equation} 
where $\delta\rho (T,H)$ is the magnetoresistance contribution.
The $T$-dependent inverse relaxation lifetime 
derived for finite field, $x\in (x_A,x_{c2})$ and
$\hbar\omega\ll \pi k_B\,T$ by replacing $\hbar\omega$ by $\pi k_B\,T$ in the 
one-electron inverse lifetime $\hbar/\tau_{el}\approx\hbar\omega\,\pi\alpha_{\tau_{el}}$ 
and averaging over the Fermi line is given by,
\begin{eqnarray}
{1\over\tau (T)} & = & {2\over\pi}\left(\int_{0}^{\pi/2}d\phi{1\over\tau_{el}}
\right)\Big\vert_{\hbar\omega=\pi k_B T} = 
{1\over\hbar A}\,\pi k_B T  \, ,
\nonumber \\
A & = & {32\over \pi^2}\sqrt{x\,x_{op}} \, , \hspace{0.25cm} x \in (x_A,x_{c2}) \, .
\label{1-over-tau}
\end{eqnarray}
The hole concentration $x_A\approx x_*/2$ is that at
which $A\approx 0.5$ becomes of order one. The normal-state
resistivity $H$-independent term $\rho (T)$ of Eq. (\ref{rho-T-TH}) then reads,
\begin{equation}
\rho (T) \approx \left({m_c^{\rho}\,d_{\parallel}\over x e^2}\right){1\over\tau (T)} 
\, ; \hspace{0.25cm} m_c^{\rho} = {\hbar^2 \pi x_*\over 2t} \, ,
\label{rho-F-tau}
\end{equation}
where $m_c^{\rho}$ is the $c$ fermion transport mass \cite{companion2}. 
Combination of Eqs. (\ref{1-over-tau}) and (\ref{rho-F-tau}) 
leads to the resistivity expression given in Eq. (\ref{rho-F}).

The method used in Ref. \cite{cuprates} to derive the one-electron inverse lifetime $1/\tau_{el}$
appearing in the inverse relaxation lifetime of Eq. (\ref{1-over-tau})
is only valid for the hole concentration range $x\in (x_A,x_{c2})$ for which
$r_{\Delta} =V^{\Delta}_{Bs1}/V_{Fc}\ll 1$. Indeed, for $x\in (x_c,x_A)$ the 
velocity ratio $r_{\Delta}$ becomes too large. As a result the 
simple form given in Ref. \cite{cuprates} for the matrix element absolute
value $\vert W_{c,s1} ({\vec{q}}^{\,h},{\vec{q}};{\vec{p}})\vert$ of the $c$ - $s1$ fermion effective 
interaction between the initial and final states is not anymore a good approximation.
Consistently with the experimental resistivity curves of Fig. \ref{fig2}, the non-linear $T$ 
dependence of the resistivity developing for approximately $x<x_A$ for a range of
low temperatures that increases upon decreasing $x$ is in part due to the
increasingly strong effects of the $s1$ fermion boundary-line velocity
anisotropy. Our method does not apply to that regime.
In turn, for the present range $x\in (x_A,x_{c2})$ the interplay of the $c$ Fermi line isotropy 
and $s1$ boundary line $d$-wave like anisotropy \cite{companion2,cuprates0} also
plays an important role. It is behind the $c$ - $s1$ fermion inelastic collisions 
leading to anisotropic one-electron scattering properties associated with
the factor $(\cos 2\phi)^2$ in the above one-electron scattering rate expression. 

The linear upper magnetic field $H_{c2}(x)$ line plotted in Refs. \cite{cuprates,cuprates0}  
for $x\in (x_0,x_{c2})$ is for $x\in (x_0,x_{c1})$ derived in Ref. \cite{cuprates0} by means of a method
introduced in Ref. \cite{duality}. Here $x_0\approx 0.013$, $x_{c1}=1/8$ and
$x_{c2}\approx 0.20$. However, for $x\in (x_1,x_{c2})$ the actual 
$H_{c2}(x)$ line may (or may not) slightly deviate to below the straight line plotted in the figure. If so, the 
hole concentration $x_{c2}\approx 0.20$ may increase to $\approx 0.21-0.22$. Fortunately, 
such a possible deviation does not change the physics discussed in this letter.

%%%%%%%%%%%%%%%%%%%%%%%%%%%%%%%%%%%%%%%%%%%%%%%%%%%%%%%%%%%%%%%%%%%%%%%%%%

%%%%%%%%%%%%%%%%%%%%%%%%%%%%%%%%%%%%%%%%%%%%%%%%%%%%%%%%%%%%%%%%%%%%%%%%%%

\end{document}